\documentclass[aip,graphicx,preprint]{revtex4-1}

\draft % marks overfull lines with a black rule on the right
\usepackage{graphicx}% Include figure files
\usepackage{dcolumn}% Align table columns on decimal point
\usepackage{bm}% bold math

\usepackage[cmex10]{amsmath}
\usepackage{amssymb}
\usepackage{gensymb}
\usepackage{subfigure}

\begin{document}

% Use the \preprint command to place your local institutional report number

% on the title page in preprint mode.

% Multiple \preprint commands are allowed.

%\preprint{}

\title{Direct calorimetric measurements of isothermal entropy change on single crystal W-type hexaferrites at the spin reorientation transition} %Title of paper

% repeat the \author .. \affiliation  etc. as needed

% \email, \thanks, \homepage, \altaffiliation all apply to the current author.

% Explanatory text should go in the []'s,

% actual e-mail address or url should go in the {}'s for \email and \homepage.

% Please use the appropriate macro for the type of information

% \affiliation command applies to all authors since the last \affiliation command.

% The \affiliation command should follow the other information.

\author{M. LoBue}
\email[]{e-mail: martino.lo-bue@satie.ens-cachan.fr}
\author{V. Loyau}
\author{F. Mazaleyrat}
\author{A. Pasko}
\affiliation{SATIE, ENS de Cachan, CNRS, UniverSud, 61 av du President Wilson, F-94230 Cachan, France}
\author{V. Basso}
\author{M. Kupferling}
\author{C. P. Sasso}
\affiliation{Istituto Nazionale di Ricerca Metrologica, Strada delle Cacce 91, 10135 Torino, Italy}

%\homepage[]{Your web page}

%\thanks{}

% Collaboration name, if desired (requires use of superscriptaddress option in \documentclass).

% \noaffiliation is required (may also be used with the \author command).

%\collaboration{}

%\noaffiliation

\date{\today}

\begin{abstract}

We report on the magnetic field induced isothermal entropy change $\Delta s(H_a, T)$ of W-type ferrite with CoZn substitution. Entropy measurement are performed by direct calorimetry. Single crystals of composition BaCo$_{0.62}$Zn$_{1.38}$Fe$_{16}$O$_{27}$, prepared by flux method, are measured at different fixed temperatures under applied field perpendicular and parallel to the $c$ axis. At $296$ K one deduces a value of K$_1=8.7\, 10^4$ J m$^{-3}$ for the first anisotropy constant, in good agreement with literature. The spin reorientation transition temperature is estimated to take place between $200$ and $220$ K.

\end{abstract}

\pacs{75.50.Gg, 73.30.Sg, 75.40.-s, 75.30.Gw}% insert suggested PACS numbers in braces on next line

\maketitle %\maketitle must follow title, authors, abstract and \pacs

% Body of paper goes here. Use proper sectioning commands.

% References should be done using the \cite, \ref, and \label commands

\section{Introduction} \label{SEC-I}

Hexagonal ferrites were intensely studied for permanent magnets and microwave absorber applications. The former related to their easy axis anisotropy configuration (e.g. in M-ferrites), the latter to an easy plane one (e.g. in Y-ferrites) \cite{Smit1959-1}. Moreover, W-type ferrites undergo spin reorientation transitions (SRT) between states of different anisotropy on varying temperature \cite{Albanese1986-1} and applied magnetic field \cite{Asti1978-1}. In spite of the lesser magnitude of the anisotropy-driven magnetocaloric effect (MCE), with respect to other materials where entropy change is associated with a change in the spontaneous magnetisation, these transitions show unique properties due to the vectorial nature of anisotropy related phenomena. The entropy change associated with SRT can be  actually achieved by changing the direction of the applied magnetic field (i.e. by a rotating field) instead of changing its amplitude. Another relevant advantage of ferrites, with respect to other alloys \cite{Basso2011-1}, is that they contain no critical rare-earth elements.

The structure of the \textbf{W} ferrites is built up by mixing the unit cells of the magnetoplumbite \textbf{M}, with chemical composition BaFe$_{12}$O$_{19}$, and that of the spinel \textbf{S}, Me$_2$Fe$_4$O$_8$. A detailed description of this structure can be found in the literature \cite{Smit1959-1}. The possibility to tune the SRT temperature by changing the proportion between the Co and Zn substitutions \cite{Albanese1986-1,Naiden1997-1} and its structural origin have been extensively studied \cite{Albanese1986-1,Paoluzi1988-1}. The system undergoes a SRT transition from a low temperature easy plane (EP) configuration to an high temperature easy axis (EA) at a temperature $T_{sr}$ that increases with Co concentration $x$. For the $x=0.62$ composition that we investigate here, the transition takes place at about $215$ K . As the magnetic field induced SRT requires the presence of a well-defined orientation of the magnetocrystalline anisotropy, the effect can be detected both in single crystal , and in aligned polycrystalline samples. In this paper we shall limit our study to single crystal, where the entropy change is expected to be greater. Comparison with polycrystalline textured samples will be presented elsewhere.

\section{Experimental} \label{SEC-II}

Single crystals of composition BaCo$_{0.62}$Zn$_{1.38}$Fe$_{16}$O$_{27}$ have been grown using a flux method \cite{Asti1978-1}. The starting mixture was prepared from BaCO$_3$, CoO, ZnO, Fe$_2$O$_3$ and Na$_2$CO$_3$ powders. The charge was packed into a platinum crucible and heated up to 1350 $^\circ$C. The melt was homogenized at this temperature for 24 h, slowly cooled down to 1000 $^\circ$C at 3 $^\circ$C/h and more rapidly to room temperature. Crystal structures were examined by PANalytical X’Pert Pro X-ray diffractometer (XRD) in Co-K$\alpha$ radiation with X'Celerator detector for rapid data acquisition.

%modified!!!!MK14/9

\begin{figure}
\includegraphics[scale=0.71]{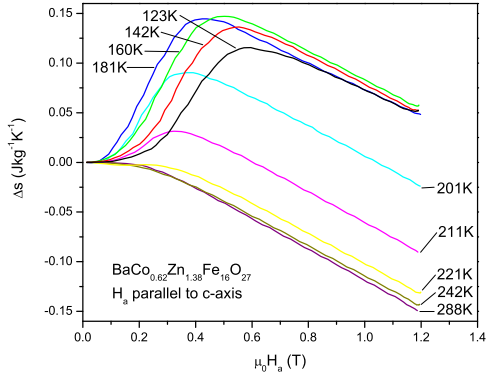}
\caption{Entropy change $\Delta s$ measured at different fixed temperatures by applying and removing a field  of $1$ T parallel to the $c$ axis.}
\label{FIG:S_Hpar}
\end{figure}

\begin{figure}
\includegraphics[scale=0.71]{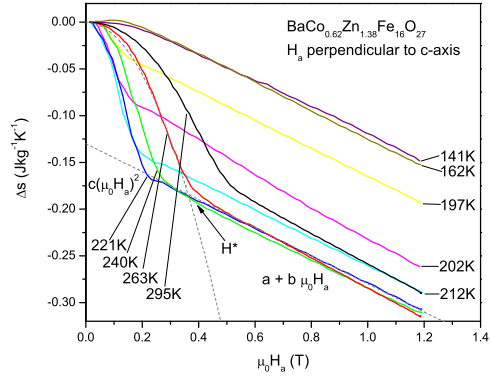}
\caption{Entropy change $\Delta s$ measured at different fixed temperatures by applying and removing a field of $1$ T perpendicular to the $c$ axis. } 
\label{FIG:S_Hperp}
\end{figure}

\begin{figure}
\includegraphics[scale=0.71]{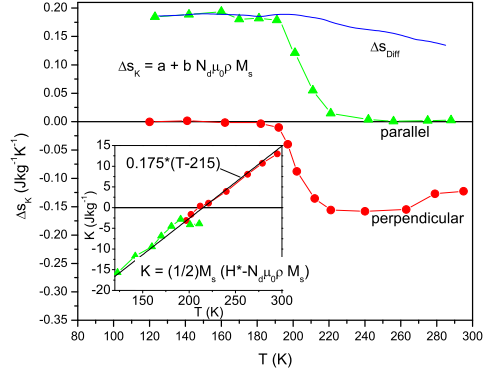}
\caption{Maximum isothermal entropy change $\Delta s_k$ in saturating parallel (green triangles) and perpendicular (red circles) fields and difference between the two curves (continuous line). Inset: equivalent anisotropy as deduced from the measured $H^*_{\parallel}(T)$ (green triangles) and $H^*_{\perp}(T)$ (red squares)  as a function of temperature}
\label{FIG:DSk}
\end{figure}

We performed measurements on a sample of mass $m=31.81$mg and of lateral size of a few millimetres. Due to the shape of the sample we estimated the demagnetising factors parallel and perpendicular to the $c$ axis to be respectively $N_{\parallel}=0.4$ and $N_{\perp}=0.3$. Isothermal entropy change has been measured using a home-made calorimetric set-up with thin film Peltier cells (Micropelt MPG D751) as heat flux sensors, working in the temperature range $77-300$ K by using an Oxford Microstat He cryostat with liquid nitrogen (a similar set-up working around room temperature is described in \cite{Basso2010-1}). The magnetic field is generated by an electromagnet (up to $2$ T maximum). The measurements are performed by keeping the temperature stable to the desired value, and meanwhile by measuring the heat flux $q_s$ during application and removal of the magnetic field (up to $1$ T) at a rate of $\mu_0 dH_a/dt=0.033$ Ts$^{-1}$. After correction of the measured heat flux for dynamic effects due to the heat transfer in the Peltier cell \cite{Basso2010-1} and an eventual offset due to non ideal isothermal conditions, the magnetic field induced entropy change is deduced by integrating $q_s$ over time. The result of the integration is represented versus the applied field $H_a$. 
In Fig.\ref{FIG:S_Hpar} and  Fig.\ref{FIG:S_Hperp} the entropy change, measured by applying the magnetic field respectively parallel and perpendicular to the $c$ axis, are shown in a range of temperatures spanning from $120$ K to $300$ K. In both cases we observe two different classes of $\Delta s(H_a)$ curves. The first is linear, $\Delta s(H_a) = b (H_{a}- N_{d} M_s)$, with negative slope $b$. The second is composed by two regimes: a quadratic one with $\Delta s(H_a) = c H_a^2$, from $H_{a}=0$ up to a field value that we shall call $H^{*}$; a linear one with $\Delta s(H_a) = \Delta s_{k} + b (H_{a}- N_d M_s)$, when $H_a  \geq H^{*}$. 
The purely linear behaviour is observed in the temperature range where the field is directed along the zero-field equilibrium configuration of the sample. The double regime is observed at temperatures where the field is applied perpendicular to the equilibrium state and therefore a field driven SRT, whose maximum entropy change is  $\Delta s_k$, takes place. From Fig. \ref{EQ:spar} and \ref{EQ:sperp} we see that $\Delta s_k$ is positive under application of $H_a$ parallel to $c$ for temperatures below about $200$ K. This corresponds to a transition from a low entropy EP configuration toward a high entropy EA one. On the contrary $\Delta s_k$ is negative when $H_a$ is\footnote{} perpendicular to $c$ and the temperature is above $200$ K. This corresponds to a transition from EA (high entropy) to EP (low entropy). 
In the main frame of Fig. \ref{FIG:DSk} the  entropy change $\Delta s_k$ (i.e. the entropy change measured at a certain temperature under application of a saturating magnetic field after subtraction of the $bH$ term) is plotted against temperature when the applied field is parallel (triangles) and perpendicular (circles) to the $c$ axis. The difference between the two curves, $\Delta s_{diff} = \Delta s_{k \parallel} - \Delta s_{k \perp}$ (continuous line) is plotted too.The presence of an intermediate temperature range, between $200$ K and $220$ K, where $\Delta s_k$ depends on temperature is apparent. The absence of detectable latent heat anomaly in DSC scans suggests that this intermediate region should be rather associated with the presence of an easy cone intermediate phase than with the phase coexistence region of a first order transition. The maximum entropy change at the reorientation is $\Delta s = 0.18$ JKg$^{-1}$K$^{-1}$. The inset of Fig. \ref{FIG:DSk} shows an effective anisotropy field $K(T)$, plotted against temperature, obtained from $H^*$ after subtraction of the demagnetising field. We shall discuss the meaning of this anisotropy in the following section.

\section{Analysis and Discussion} \label{SEC:III}

The system free energy with anisotropy term developed till the third order writes: 

\begin{eqnarray}\label{EQ:GL}
 G_L(\theta, \textbf{H}, T) = G_0(T) + K_1 sin^2 \theta + K_2 sin^4 \theta + \nonumber \\   
+ K_3 sin^6 \theta - \mu_0 M_s(H_{\parallel} cos \theta + H_{\perp} sin \theta),
\end{eqnarray}
 
where $\theta$ is the angle between the magnetisation $\textbf{M}$ and the $c$
axis, $H_{\parallel}$ and $H_{\perp}$ are the components of $\textbf{H}$ respectively parallel and perpendicular to $c$, $K_1$, $K_2$, $K_3$ are the anisotropy constants, and $G_0(T)$ is a term depending on temperature only. The entropy of the system under a saturating field directed along the $c$ axis ($H_{a \parallel} \geq H^*_{\parallel}$) is: 

\begin{equation}\label{EQ:spar}
s_{\parallel} = - \frac{dG_0}{dT} + \mu_0 \frac{dM_s}{dT} \left( H_{a \parallel} -  N_{\parallel} M_s \right)
\end{equation}

When the field is perpendicular to $c$ ($H_{a \perp} \geq H^*_{\perp}$) we have: 

\begin{eqnarray}\label{EQ:sperp}
 s_{\perp} = - \frac{dG_0}{dT} - \frac{dK_1}{dT} - \frac{dK_2}{dT}- \frac{dK_3}{dT} + \nonumber \\ 
+ \mu_0 \frac{dM_s}{dT } \left( H_{a \perp} - N_{\perp} M_s \right)
\end{eqnarray}

Now, whatever the initial state $s(0)$ (i.e. the $H_a =0$ entropy at a fixed temperature) from which the $\Delta s(H_a)$ measurements are performed, $s(0)$ will be nothing but a function of $T$. So under saturating applied field (i.e. when $H_a \geq H^*$) from Eq. (\ref{EQ:spar}) and (\ref{EQ:sperp}) we can deduce that $b= \mu_0 dM_s/dT$. By fitting the linear part in the field region where $H_a \geq H^*$ we obtain $dM_s/dT = 802.5$ Am$^{-1}$, in good agreement with values from the literature \cite{Paoluzi1988-1}. Using the $dM_s/dT$ value and Eq. (\ref{EQ:spar}) and (\ref{EQ:sperp}) we can subtract from $\Delta s$ the contribution of paraprocesses and plot the SRT contribution to the entropy change  $\Delta s_k$ in parallel and perpendicular $H_a$. From Eq. (\ref{EQ:spar}) and (\ref{EQ:sperp}) their difference is $\Delta s_{diff} = \Delta s_{k \parallel} - \Delta s_{k \perp} = dK_1/dT + dK_2/dT +  dK_3/dT$. From the data shown in Fig. \ref{FIG:DSk} we find that $\Delta s_{diff} = 0.2 - 0.6\, 10^{-3} (T-205)$ with a slight negative slope. 

Now, let us discuss the fields $H^*_{\parallel}$ and $H^*_{\perp}$, representing the $H_a$ value where $\Delta s(H_a)$ pass from the quadratic regime to the linear one. During a uniform rotation of the magnetization driven by $H_a$, the system reaches saturation when the applied field overcomes both the demagnetising and the anisotropy field. Therefore, for a rotation from the easy plane towards the $c$ axis (i.e. the parallel field case) we have $H^*_{\parallel} = H_{k1} + N_{\parallel} M_s$ where $H_{k1} = 2 \left| K_1 \right| / \mu_0 M_s$, whereas when magnetisation rotates from the $c$ axis to the plane, $H^*_{\perp} = H_{k2} + N_{\perp} M_s$ where $H_{k2} = 2 \left| K_1 + 2 K_2 + 3 K_3  \right| / \mu_0 M_s$. It is worth recalling that, in the case where $K_2 = K_3 = 0$, the quadratic expression $\Delta s = cH^2_a$ for $H_a<H^*$ can be deduced rigorously from expression (\ref{EQ:GL}) \cite{Basso2011-1}. When $K_2$ and $K_3$ can not be neglected, stable zero-field easy-cone configuration may exist, and a simple expression for $\Delta s(H_a)$ (when $H_a<H^*$) can be hardly worked out. Notwithstanding this limitation our measured $\Delta s(H_a)$ curves actually show a quadratic and a linear regime as apparent from Fig. \ref{FIG:S_Hpar} and \ref{FIG:S_Hperp}. So we must just keep in mind that identification of $H^*$ is theoretically founded near room temperature, whereas it represents just a phenomenological fitting procedure for lower temperatures. From $H^*$, by subtracting the demagnetising field, we can calculate the value of $H_{k1}$ and $H_{k2}$ and therefore we can respectively work out two anisotropies: $K_{EP} = K_1 + 2 K_2 + 3 K_3$ (red circles in the inset of Fig. \ref{FIG:DSk})  and $K_1$ (green triangles in the inset of Fig. \ref{FIG:DSk}). As higher order anisotropy constants are expected to vanish at room temperature, from the value of $K_{EP}$ at $296$ K we can directly estimate $K_1 = 8.7\, 10^4$ Jm$^{-3}$, in good agreement with the literature \cite{Asti1978-1}. By fitting $K_{EP}(T)$ and $K_1(T)$ together with a linear function we obtain: $K(T) = 0.175 \left(T - 215 \right)$ JKg$^{-1}$. This result is rather intriguing as, from magnetic measurements, it is generally argued that at low temperature higher order anisotropy constants should not be negligible \cite{Asti1978-1,Naiden1990-1}. However, values and sign of the anisotropy constants deduced from magnetic and torque measurements are often conflicting \cite{Asti1978-1,Naiden1990-1,Smit1959-1}. Indeed, the calorimetric measurement we present here, can be rather well described by using just one equivalent anisotropy constant that changes its sign at $215$ K, inside the temperature interval where the SRT takes place. Discrepancies between the values of $K$ deduced from $H^*_{\perp}$ (red circles) and from $H^*_{\parallel}$ (green triangles) are apparent around $200$ K. Moreover, the slope of $K(T)$ is constant, $dK/dT = 0.175$ JKg$^{-1}$K$^{-1}$, whereas from $\Delta s_{diff}(T)$ we found that $K_1+K_2+K_3$ presents a slight negative slope. This can be ascribed as due to the contribution of higher order constant. Further investigation on this issue will be focused on temperature dependence of magnetisation curves and on more detailed theoretical calculations.

Concluding: we presented a detailed experimental investigation on SRT associated entropy change in W-type hexaferrite covering a wide temperature range. The data, the first in our knowledge obtained by direct calorimetry, allow to identify: the temperature interval where the transition takes place, the maximum entropy change associated with SRT, and can be described using a single anisotropy constant $K$ model where $K$ changes sign at the transition temperature. Further investigation will be devoted to the role of higher anisotropy constants.

\begin{acknowledgments}

% Put your acknowledgments here.
The research leading to this results has received funding from the European Community's $7^{th}$ Framework Programme under Grant Agreement No. 214864 (project SSEEC).

\end{acknowledgments}

\newpage

%\begin{figure}
%\includegraphics[scale=1.5]{S_Hpar.eps}
%\caption{Entropy change $\Delta s$ measured at different fixed temperatures by applying and removing a field  of $1$ T parallel to the $c$ axis.}
%\label{FIG:S_Hpar}
%\end{figure}

%\begin{figure}
%\includegraphics[scale=1.5]{S_Hperp.eps}
%\caption{Entropy change $\Delta s$ measured at different fixed temperatures by applying and removing a field of $1$ T perpendicular to the $c$ axis. } 
%\label{FIG:S_Hperp}
%\end{figure}

%\begin{figure}
%\includegraphics[scale=1.5]{deltask.eps}
%\caption{Maximum isothermal entropy change $\Delta s_k$ in saturating parallel (green triangles) and perpendicular (red circles) fields and difference between the two curves (continuous line). Inset: equivalent anisotropy as deduced from the measured $H^*_{\parallel}(T)$ (green triangles) and $H^*_{\perp}(T)$ (red squares)  as a function of temperature}
%\label{FIG:DSk}
%\end{figure}

\end{document}